\documentclass[review,12pt]{elsarticle}

\usepackage{amssymb}
\usepackage{amsthm}
\usepackage{amsmath}
\usepackage{mathrsfs}
\usepackage{graphicx}
\usepackage{epstopdf}
\usepackage{float}
\usepackage{caption}
\usepackage{soul}
\usepackage{color,soul} 
\usepackage{color} 
\usepackage{subcaption}
\usepackage{bm}
\usepackage{bbm}
\usepackage{mathrsfs}
\usepackage{soul}
\usepackage{hyperref} 
\usepackage{cleveref}
\usepackage[margin=2.3cm]{geometry}
\biboptions{super,sort&compress}

\hypersetup{
    colorlinks=true,
    linkcolor=blue,
    filecolor=magenta,      
    urlcolor=cyan,
    pdftitle={Overleaf Example},
    pdfpagemode=FullScreen,
    } 

\journal{ }

\makeatletter
\def\@author#1{\g@addto@macro\elsauthors{\normalsize%
    \def\baselinestretch{1}%
    \upshape\authorsep#1\unskip\textsuperscript{%
      \ifx\@fnmark\@empty\else\unskip\sep\@fnmark\let\sep=,\fi
      \ifx\@corref\@empty\else\unskip\sep\@corref\let\sep=,\fi
      }%
    \def\authorsep{\unskip,\space}%
    \global\let\@fnmark\@empty
    \global\let\@corref\@empty  
    \global\let\sep\@empty}%
    \@eadauthor={#1}
}
\makeatother

\makeatletter
\def\thickhline{%
  \noalign{\ifnum0=`}\fi\hrule \@height \thickarrayrulewidth \futurelet
   \reserved@a\@xthickhline}
\def\@xthickhline{\ifx\reserved@a\thickhline
               \vskip\doublerulesep
               \vskip-\thickarrayrulewidth
             \fi
      \ifnum0=`{\fi}}
\makeatother

\newlength{\thickarrayrulewidth}
\begin{document}

\begin{frontmatter}



\title{Modelling fatigue crack growth in Shape Memory Alloys}


\author{Marlini Simoes\fnref{Cam,ESA}}

\author{Christopher Braithwaite\fnref{Cam}}

\author{Advenit Makaya\fnref{ESA}}

\author{Emilio Mart\'{\i}nez-Pa\~neda\corref{cor1}\fnref{IC}}
\ead{e.martinez-paneda@imperial.ac.uk}

\address[Cam]{Cavendish Laboratory, University of Cambridge, Cambridge CB3 0HE, UK}

\address[ESA]{ESA-ESTEC, Keplerlaan 1, 2200, AG Noordwijk-ZH, Netherlands}

\address[IC]{Department of Civil and Environmental Engineering, Imperial College London, London SW7 2AZ, UK}

\cortext[cor1]{Corresponding author.}

\begin{abstract}
We present a phase field-based framework for modelling fatigue damage in Shape Memory Alloys (SMAs). The model combines, for the first time: (i) a generalised phase field description of fracture, incorporating multiple phase field formulations, (ii) a constitutive model for SMAs, based on a Drucker-Prager form of the transformation surface, and (iii) a fatigue degradation function, with damage driven by both elastic and transformation strains. The theoretical framework is numerically implemented, and the resulting linearised system is solved using a robust monolithic scheme, based on quasi-Newton methods. Several paradigmatic boundary value problems are addressed to gain insight into the role of transformation stresses, stress-strain hysteresis and temperature. Namely, we compute $\Delta \varepsilon-N$ curves, quantify Paris law parameters and predict fatigue crack growth rates in several geometries. In addition, the potential of the model for solving large-scale problems is demonstrated by simulating the fatigue failure of a 3D lattice structure.
\end{abstract}


\end{frontmatter}


\section{Introduction}
\label{Introduction}

\begin{figure}[H]
  \makebox[\textwidth][c]{\includegraphics[width=0.9\textwidth]{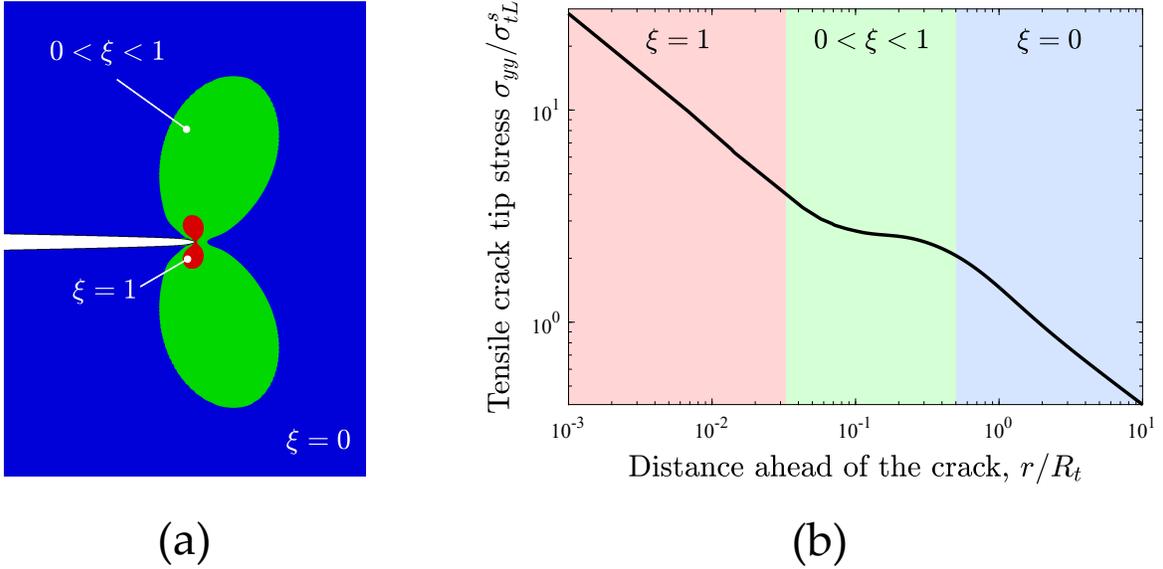}}%
  \caption{Crack tip mechanics of SMAs under small scale transformation zone conditions: (a) crack tip contours, and (b) tensile stress distribution ahead of the crack, in a log-log plot. Three regions are differentiated, based on the magnitude of the martensitic volume fraction $\xi$. The tensile stress distribution is normalised by the transformation stress threshold $\sigma_{tL}^s$ and the distance ahead of the crack $r$ is normalised by an Irwin-like estimation of the transformation zone length: $R_t=(1/3\pi)(K_I/\sigma_{tL}^s)^2$, where $K_I$ is the stress intensity factor.}
  \label{fig:Stationary}
\end{figure}

Shape Memory Alloys (SMAs) have received significant attention due to their remarkable properties \cite{Lagoudas2008}. Owing to their capacity to undergo a thermal and/or stress induced solid-to-solid phase transformation, SMAs exhibit the largest reversible strains of all crystalline materials - of up to 10\%, one order of magnitude higher than traditional alloys. The ability of SMAs to recover their original shape and display a recoverable non-linear response under very large strains has fostered their application in a wide range of sectors, from biomedical devices to aerospace components \cite{Robertson2007,Hartl2007}. In many of these applications, structural integrity is paramount. Experiments and theoretical modelling have shown that there is a toughening effect associated with phase transformation \cite{Baxevanis2015,Gollerthan2009,Baxevanis2013}. Transformation strains resulting from the phase change provide a source of energy dissipation that enhances crack growth resistance \cite{Freed2007,CMAME2021}. The yield stress of SMAs is typically much larger than the threshold for stress induced transformation and, as a result, a transformation zone develops in the vicinity of the crack tip. As shown in Fig. \ref{fig:Stationary} for the case of a NiTi alloy (arguably the most widely exploited class of SMAs), the distribution of crack tip stresses shows three regions: (1) an outer austenitic region ($\xi=0$), (2) an intermediate austenite-martensite transformation region ($0 < \xi < 1$), and (3) an inner martensitic region ($\xi=1$) \cite{Maletta2013,CMAME2021}. These features govern the fracture and fatigue response of SMAs \cite{Wilkes2000,Baxevanis2014,Haghgouyan2019}.

The fatigue behaviour of SMAs has been the focus of much attention as the most successful applications of SMAs involve significant cyclic mechanical motion; e.g., actuator springs or medical devices such as stents \cite{Hornbogen2002,Eggeler2004,Hornbogen2004,Lagoudas2009,Robertson2012,Ghafoori2017}. However, the vast majority of the literature in the area of SMA fatigue is experimental, and there is a need for modelling endeavours that can provide further insight, optimise design, and map safe regimes of operation. While efforts have been made to model the constitutive behaviour of SMAs under cyclic loading \cite{Wang2019,MohammadHashemi2019,Xiao2020}, and to quantify the stationary stress fields ahead of cracks and notches \cite{Wang2007a,Wang2010}, the study of sub-critical crack growth remains an elusive goal. This is despite its importance for defect-tolerant design approaches. Very recently, Simoes and Mart\'{\i}nez-Pa\~neda \cite{CMAME2021} presented a phase field formulation for modelling crack growth in SMAs. They formulated a generalised phase field theory and combined it with the unified SMA model by Lagoudas \textit{et al.} \cite{Lagoudas1996}. Phase field models can readily capture cracking phenomena of arbitrary complexity and have been recently extended to fatigue damage (see, e.g. \cite{Wu2020,Materials2021,Carrara2020,Golahmar2022}, and Refs. therein). 

In this work, we develop a new model for fatigue crack growth in SMAs. The constitutive SMA model by Auricchio and co-workers \cite{Auricchio1997} is combined with a phase field formulation for fracture and fatigue, where the damage driving force involves both elastic and transformation strains. The model is general and aims at capturing a wide range of material behaviour, thus accommodating both \texttt{AT1} \cite{Pham2011} and \texttt{AT2} \cite{Bourdin2000} phase field theories. Numerical robustness and efficiency is achieved by using a quasi-Newton based monolithic solution scheme \cite{Wu2020a,TAFM2020}. Several paradigmatic boundary value problems are investigated to gain insight into the fatigue response of SMAs. Firstly, we produce \textit{Virtual} $\Delta \varepsilon-N$ curves and assess the influence of temperature and stress-strain hysteresis. Secondly, crack extension is predicted in a single-edged notched tension specimen considering different definitions of the crack density function. Thirdly, we obtain $da/dN$ vs $\Delta K$ curves and predict the associated Paris law parameters for different material modelling choices. Finally, we investigate the evolution of fatigue cracks in a 3D SMA lattice structure undergoing cyclic loading.

The remainder of this manuscript is organised as follows. In Section \ref{Sec:Theory}, we present the theoretical framework for our phase field fatigue SMA model. Details of the numerical implementation are provided in Section \ref{Sec:NumModel}. Section \ref{Sec:FEMresults} is dedicated to showcasing the predictions of the model and discussing the interpretation of the finite element results presented in the context of the literature. Finally, concluding remarks are given in Section \ref{Sec:ConcludingRemarks}. 

\section{A phase field fatigue formulation for Shape Memory Alloys}
\label{Sec:Theory}

Here, the theory underlying our phase field-based model for fatigue in SMAs is presented. First, in Section \ref{Sec:PhaseFieldFracture}, we describe the phase field regularisation of fracture. Then, in Section \ref{Sec:PhaseFieldFatigue}, the extension to fatigue and the variational formulation are presented. Finally, Section \ref{sec:ConstitutiveSMA} provides a brief outline of the constitutive model employed to capture the mechanical response of Shape Memory Alloys (SMAs). For simplicity, small strains and isothermal behaviour is assumed; we will assess the influence of temperature by conducting numerical experiments at different temperatures but disregard thermal loads. Plastic deformations are also considered a secondary effect and neglected for simplicity, as the yield stress of SMAs is typically much larger than the transformation stress.

\subsection{A phase field description of fracture}
\label{Sec:PhaseFieldFracture}

Phase field methods have become very popular in the modelling of a wide range of interfacial problems, from microstructural evolution \cite{Provatas2011} to corrosion \cite{JMPS2021}. By using an auxiliary variable, the phase field $\phi$, a given interface can be smeared over the computational domain. The interface is no longer sharp and of zero thickness, but smooth and of a finite thickness, as governed by a length scale parameter. This paradigm greatly facilitates the computational treatment of evolving interfaces, as the interface equation is defined in the entire domain (i.e., no special treatment is needed) and topological changes such as divisions or merging of interfaces are naturally captured. In fracture mechanics, the phase field describes the location of the crack (solid-crack interface), taking (e.g.) a value of $\phi=0$ in intact material points and of $\phi=1$ in fully cracked material points, and varying smoothly in-between (akin to a damage variable). Phase field fracture methods have opened new modelling horizons, enabling the prediction of complex cracking phenomena such as crack branching, deflection, nucleation and merging in arbitrary geometries and dimensions \cite{Borden2012,TAFM2020c,Zhou2019d,Peng2021}. Applications have soared and now include the fracture of functionally graded materials \cite{CPB2019,Kumar2021}, composites \cite{Quintanas-Corominas2019,CST2021,Bui2021}, rock-like materials \cite{Zhou2018,Aldakheel2021}, ductile \cite{Borden2016,Alessi2018,Shishvan2021a} and embrittled \cite{CMAME2018,Anand2019,JMPS2020} metals, and natural materials \cite{Yin2019,Mandal2020a}. The evolution of the phase field $\phi$ is generally defined based on Griffith's \cite{Griffith1920} energy balance and the thermodynamics of fracture \cite{Bourdin2000,Bourdin2008}. Thus, crack growth is driven by the competition between the energy released by the solid and the energy required to create two new surfaces, often referred to as $G_c$, the critical energy release rate or the material toughness. Accordingly, the diffuse phase field representation introduces the following approximation of the fracture energy over a discontinuous surface $\Gamma$:
\begin{equation}
    \Phi=\int_{\Gamma} G_c \, \text{d}S \approx \int_\Omega  G_c \left[ \frac{1}{4c_w \ell} \Big( w (\phi) + \ell^2 |\nabla \phi|^2 \Big) \right] \, \text{d}V, \hspace{0.8cm} \text{for } \ell\rightarrow 0^+.
\end{equation}

\noindent Here, $\ell$ is the phase field length scale, $w (\phi)$ is the geometric crack function and $c_w$ is a scaling constant, defined as $c_w=\int_0^1 \sqrt{w(\zeta)} \text{d} \zeta$. The choice of $w (\phi)$ defines the damage constitutive behavior. In this work, we aim at providing a generalised framework by incorporating both the so-called \texttt{AT1} \cite{Pham2011} and \texttt{AT2} \cite{Bourdin2000} phase field models. The \texttt{AT2} formulation is recovered for $w(\phi)=\phi^2$ ($c_w=1/2$) while in the \texttt{AT1} model $w(\phi)=\phi$ ($c_w=2/3$). Thus, the main difference between the \texttt{AT1} and the \texttt{AT2} models is that the latter does not have a damage threshold (as $w'(0)=0$), while the \texttt{AT1} formulation exhibits a linear elastic regime prior to the onset of damage.

It has been recently emphasised that for a finite value of $\ell$, variational phase field fracture models exhibit a finite strength, which enables capturing the crack size effect \cite{Tanne2018,PTRSA2021}. The critical failure stress and the critical failure strain can be related to the elastic and fracture properties by solving the homogeneous 1D problem ($\nabla \phi=0$), giving \cite{PTRSA2021}:
\begin{align}\label{eq:AT1scec}
    & \texttt{AT1}: \,\,\, \sigma_c = \sqrt{\frac{3EG_c}{8 \ell}} \, , \,\,\,\,\,\,\,\,\,\,\, \varepsilon_c =  \sqrt{\frac{3G_c}{8\ell E}} \, . \\
    &  \texttt{AT2}: \,\,\, \sigma_c = \sqrt{\frac{27EG_c}{256 \ell}}  \, , \,\,\,\,\,\,\,\,\,\, \varepsilon_c =  \sqrt{\frac{G_c}{3\ell E}} \, . \label{eq:AT2scec}
\end{align}


\subsection{Variational phase field fatigue}
\label{Sec:PhaseFieldFatigue}

Let us now formulate the variational problem, incorporating both bulk $\Psi^b$ and surface $\Psi^s$ energies, and introduce fatigue damage into the formulation. The latter is achieved by defining a fatigue degradation function $f(\bar{\alpha})$, where $\bar{\alpha}$ is a cumulative history variable. Accordingly, the total potential energy is given by:
\begin{equation}\label{Eq:Piphi}
     \Psi =  \Psi^b \left( \bm{\varepsilon}, \xi, \phi \right) + \Psi^s \left( \phi,  \bar{\alpha} \right) =   \int_\Omega \bigg\{ \left( 1 - \phi \right)^2  \psi \left( \bm{\varepsilon},  \xi \right)  +  f \left( \bar{\alpha} \right) \frac{G_c }{4 c_w \ell} \left(   w (\phi) + \ell^2 |\nabla \phi|^2 \right) \bigg\} \, \text{d} V \, ,
\end{equation}

\noindent where $\psi$ is the strain energy density and $\bm{\varepsilon}$ is the strain tensor. Both can be divided into their elastic and transformation parts, such that the total strain energy density $\psi$ is given by:
\begin{equation}\label{eq:PsiTotal}
    \psi \left(\bm{\varepsilon},  \xi \right) = \int_0^t \left( \bm{\sigma} : \dot{\bm{\varepsilon}}^e \right) \, \text{d}t + \int_0^t \left( \bm{\sigma} : \dot{\bm{\varepsilon}}^{t} \right) \, \text{d}t \, ,
\end{equation}

\noindent where $\bm{\varepsilon}^e$ and $\bm{\varepsilon}^t$ respectively denote the elastic and transformation strain tensors. Since the fracture driving force is the total strain energy density, both elastic and transformation strains are assumed to contribute to material damage.

We proceed now to define the fatigue degradation function $f(\bar{\alpha})$. Here, we follow the work by Carrara \textit{et al.} \cite{Carrara2020} and adopt the following asymptotic function:
\begin{equation}\label{eq:fdeg1}
   {f}(\bar{\alpha}(t))=
   \begin{cases}
            1 &         \text{if } \Bar{\alpha}(t)\leq\alpha_T\\
            \left(\frac{2\alpha_T}{\Bar{\alpha}(t)+\alpha_T}\right)^2 &         \text{if } \Bar{\alpha}(t)\geq\alpha_T
   \end{cases}
\end{equation}

\noindent where $\alpha_T$ is a fatigue threshold parameter, defined as: $\alpha_T=G_c/(12\ell)$ \cite{Carrara2020}. The fatigue history variable $\bar{\alpha}$ is computed as,
\begin{equation}\label{eq:alpha_bar1}
   \Bar{\alpha}(t)=\int_0^t H(\alpha\dot{\alpha})|\dot{\alpha}|\text{ d}t \, ,
\end{equation}

\noindent where $H(\alpha\dot{\alpha})$ is the Heaviside function, introduced to ensure that $\bar{\alpha}$ only increases during loading. Finally, consistent with our definition of the fracture driving force, the fatigue history variable is defined as:
\begin{equation}
    \alpha = \left( 1 - \phi \right)^2 \psi \, .
\end{equation}

\subsection{Constitutive behaviour of SMAs}
\label{sec:ConstitutiveSMA}

The constitutive response of the phase transforming solid is based on the material model for SMAs developed Auricchio and co-workers \cite{Auricchio1997,Auricchio1997b}. Accordingly, the elastic properties (Young's modulus $E$ and Poisson's ratio $\nu$) are derived using the rule of mixtures from the corresponding austenitic and martensitic properties, where $\xi$ is the volume fraction of martensite; i.e.:
\begin{equation}
    E=E_A+\xi \left( E_M - E_A \right) \, ,
\end{equation}
\begin{equation}
    \nu = \nu_A + \xi \left( \nu_M - \nu_A \right) \, .
\end{equation}

The evolution of the transformation strains is estimated using a Drucker-Prager type of loading function. The increment in transformation strain is given by the following flow rule:
\begin{equation}
    \Delta \bm{\varepsilon}^t = \Delta \xi \frac{\partial G^t}{\partial \bm{\sigma}} \, .
\end{equation}

\noindent where $G^t$ is the transformation flow potential. Both $G^t$ and the transformation surface $F^t$ are assumed to follow a Drucker-Prager form:
\begin{equation}
    G^t = \sqrt{\frac{3}{2} \bm{\sigma}':\bm{\sigma}'} + \frac{1}{3} \text{tr} \left( \bm{\sigma} \right) \tan \varphi \, .
\end{equation}
\begin{equation}
    F^t = \sqrt{\frac{3}{2} \bm{\sigma}':\bm{\sigma}'} + \frac{1}{3} \text{tr} \left( \bm{\sigma} \right) \tan \beta \, .
\end{equation}

\noindent Here, the angles $\varphi$ and $\beta$ are material constants estimated from the tensile and compressive transformation stress levels, the uniaxial transformation strain, and the volumetric transformation strain \cite{Auricchio1997}. This ensures that the pressure dependency of phase transformations is captured. We emphasise that other constitutive SMA choices can be coupled with the phase field fatigue descriptions of fracture and fatigue presented in Sections \ref{Sec:PhaseFieldFracture} and \ref{Sec:PhaseFieldFatigue}, respectively.

\section{Numerical implementation}
\label{Sec:NumModel}

Details of the numerical implementation are provided here, starting with the enforcement of damage irreversibility (Section \ref{Sec:Irreversibility}), followed by the finite element discretisation and formulation of the residuals and stiffness matrices (Section \ref{Sec:FEdiscretisation}), the quasi-Newton solution scheme employed (Section \ref{Sec:SolutionScheme}), and the particularities of the implementation in the commercial finite element package \texttt{ABAQUS} through a user-defined \texttt{UELMAT} subroutine (Section \ref{Sec:ABAQUSdetails}). 

\subsection{Damage irreversibility}
\label{Sec:Irreversibility}

A history variable field $\mathcal{H}$ is introduced to ensure damage irreversibility, $\phi_{n+1} \geq \phi_{n}$. Thus, the history field must satisfy the Kuhn-Tucker conditions
\begin{equation}
    \psi - \mathcal{H} \leq 0 \text{,} \hspace{7mm} \dot{\mathcal{H}} \geq 0 \text{,} \hspace{7mm} \dot{\mathcal{H}}(\psi -\mathcal{H})=0
    \centering
\end{equation}

Accordingly, for a current time $t$, the history field can be defined as
\begin{equation}
     \mathcal{H} = \max_{\tau \in[0,t]}\psi ( \tau) \, . 
\end{equation}

Also, one should note that the \texttt{AT1} phase field formulation does not inherently ensure that the lower bound on the phase field is enforced. If no measures are taken, the phase field can become negative for strains below the critical one, $\varepsilon_c$, Eq. (\ref{eq:AT1scec}). To prevent this, we define the following threshold value for the history field:
\begin{equation}\label{eq:HminB}
  \mathcal{H}_{min}^{\texttt{AT1}} = \frac{3G_c}{16\ell} 
\end{equation}

\subsection{Finite element discretisation}
\label{Sec:FEdiscretisation}

We shall now formulate the two-field weak form and subsequently derive the stiffness matrices and residuals applying a finite element discretisation. Thus, consider the total potential energy of the solid, Eq. (\ref{Eq:Piphi}). In the absence of body forces and external tractions, the first variation of (\ref{Eq:Piphi}) with respect to $\bm{\varepsilon}$ and $\phi$, gives
\begin{align} \label{eq:weak1}
    \int_\Omega \left[ \left( 1 - \phi \right)^2 \bm{\sigma}: \delta \bm{\varepsilon} \right] \text{d} V &= 0 \, ,  \\
     \label{eq:weak2}
  \int_{\Omega} \left[ -2(1-\phi)\delta \phi \, \mathcal{H} +
        f \left( \bar{\alpha} \right) G_c  \left( \dfrac{\phi}{\ell} \delta \phi
        + \ell\nabla \phi \cdot \nabla \delta \phi \right) \right]  \, \mathrm{d}V &= 0   \, , 
\end{align}

\noindent where we have made use of the history field concept. Now, adopting Voigt notation, consider the following finite element interpolation for the nodal variables: the displacement vector $\mathbf{u}$ and the phase field $\phi$,
\begin{equation}
    \mathbf{u}=\sum_{i=1}^{m} \bm{N}_{i}^{\mathbf{u}} \mathbf{u}_{i} \hspace{7mm} \text{and} \hspace{7mm} \phi=\sum_{i=1}^{m} N_{i} \phi_{i} 
    \centering
\end{equation}

\noindent where $n$ is the number of nodes and $\bm{N}_{i}$ denotes the shape function matrix, a diagonal matrix with $N_i$ in the diagonal terms. Consequently, the corresponding derivatives can be discretised making use of the strain-displacement matrices $\bm{B}_{i}^{\mathbf{u}}$ and $\bm{B}_{i}^{\phi}$ as follows:
\begin{equation}
    \bm{\varepsilon}=\sum_{i=1}^{m} \bm{B}_{i}^{\mathbf{u}} \mathbf{u}_{i} \hspace{7mm} \text{and} \hspace{7mm} \nabla \phi=\sum_{i=1}^{m} \bm{B}_{i}^{\phi} \phi_{i} \, .
    \centering
\end{equation}

Making use of this discretisation and considering that (\ref{eq:weak1})-(\ref{eq:weak2}) must hold for arbitrary values of the primal kinematic variables, the residuals are derived as:
\begin{align}
    \bm{r}_{i}^{\mathbf{u}} = \int_{\Omega} \left\{  \left[(1-\phi)^{2} + \kappa \right] {(\bm{B}_{i}^{\mathbf{u}})}^{T} \bm{\sigma} \right\} \, \text{d}V & \\ 
    r_{i}^{\phi}= \int_{\Omega} \left\{ -2(1-\phi) N_{i} \, \mathcal{H} +
   \frac{G_c}{2c_w \ell} \left[ \frac{w'(\phi)}{2} N_{i} 
    + \ell^2 {(\bm{B}_{i}^{\phi})}^{T} \nabla \phi \right] \right\} \, \text{d}V  &  
\end{align}

\noindent where $\kappa$ is a small positive constant introduced to keep the system of equations well-conditioned. As commonly done in the literature (see, e.g., \cite{CMAME2018}), a value of $\kappa= 1 \times 10^{-7}$ is adopted throughout this manuscript. Finally, the tangent stiffness matrices are derived as:
\begin{align}
  \bm{K}_{ij}^{\mathbf{u} \mathbf{u}} = \frac{\partial \bm{r}_{i}^{\mathbf{u}} }{\partial \mathbf{u}_{j} } = 
        \int_{\Omega} \left\{ \left[ (1-\phi)^2+\kappa \right] {(\bm{B}_{i}^{\mathbf{u}})}^{T} \mathcal{\bm{C}} \, \bm{B}_{j}^{\mathbf{u}} \right\} \, \text{d}V & \\
        \bm{K}_{ij}^{\phi \phi} = \frac{\partial r_{i}^{\phi} }{ \partial \phi_{j} } =  \int_{\Omega} \left\{ \left( 2 \mathcal{H} + \frac{G_{c}}{4 c_w \ell} w'' (\phi) \right) N_{i} N_{j} + \frac{G_{c} \ell}{2c_w} \, {(\bm{B}_{i}^{\phi})}^{T} (\bm{B}_{j}^{\phi}) \right\} \, \text{d}V &
\end{align}

\noindent where $\mathcal{\bm{C}}$ is the material Jacobian.

\subsection{Solution scheme}
\label{Sec:SolutionScheme}

The resulting system of equations is solved in a monolithic manner, using a quasi-Newton approach, which requires the stiffness matrix to be symmetric and positive-definite. Accordingly, the non-diagonal terms of the stiffness matrix are taken to be equal to zero and the resulting system reads:
\begin{equation}\label{Eq:GlobalElementSystem}
    {\begin{Bmatrix}
        \textbf{u}\\[0.3em] \bm{\phi}
    \end{Bmatrix}}_{t+\Delta t} = 
    {\begin{Bmatrix}
        \textbf{u}\\[0.3em] \bm{\phi}
    \end{Bmatrix}}_{t} -
    {\begin{bmatrix}
        \mathbf{K}^{\mathbf{u}\mathbf{u}} & 0 \\[0.3em] 
        0 & \mathbf{K}^{\phi\phi}
    \end{bmatrix}}_{t}^{-1}
    {\begin{Bmatrix}
        \textbf{r}^{\textbf{u}}\\[0.3em] \textbf{r}^{\phi}
    \end{Bmatrix}}_{t} \, .
    \centering
\end{equation}

A global iterative scheme is adopted to obtain the solutions for which $\bm{r}^{\mathbf{u}}=\bm{0}$ and $\textbf{r}^{\phi}=\bm{0}$. The total potential energy functional is non-convex with respect to $\mathbf{u}$ and $\phi$ but quasi-Newton methods have shown to be very robust when dealing with non-convex minimisation problems \cite{Wu2020a,TAFM2020}. Specifically, the so-called BFGS (Broyden-Fletcher-Goldfarb-Shanno) algorithm is used \cite{Matthies1979}. Thus, if the linearised system is described as:
\begin{equation}
    \tilde{\mathbf{K}}\Delta\mathbf{z}= \Delta\mathbf{r} \,\,\,\,\,\,\,\,\, \text{with} \,\,\,\,\,\,\,\,\,  \mathbf{z}=\begin{Bmatrix} \mathbf{u}\\[0.3em] \bm{\phi}\end{Bmatrix} \, ,
\end{equation}

\noindent then the approximated stiffness matrix is defined as follows:
\begin{equation}
    \tilde{\mathbf{K}} = \tilde{\mathbf{K}}_t - \dfrac{(\tilde{\mathbf{K}}_t\Delta\mathbf{z})(\tilde{\mathbf{K}}_t)\Delta\mathbf{z})^T}{\Delta\mathbf{z}\tilde{\mathbf{K}}_t\Delta\mathbf{z}}+\dfrac{\Delta\mathbf{r}\Delta\mathbf{r}^T}{\Delta\mathbf{z}^T\Delta\mathbf{r}} \, .
\end{equation}

Since the SMA constitutive model is implemented using an implicit integration scheme, unconditional stability is guaranteed, speeding up calculations by several orders of magnitude relative to staggered solution schemes \cite{TAFM2020}.

\subsection{Details of the \texttt{ABAQUS} implementation}
\label{Sec:ABAQUSdetails}

A \texttt{UELMAT} subroutine is developed to implement the theory described in Section \ref{Sec:Theory} into the commercial finite element package \texttt{ABAQUS}. The \texttt{UELMAT}, like a user element (\texttt{UEL}) subroutine, requires defining the element residual and stiffness matrices; \texttt{ABAQUS} is used only to assemble the global matrices and solve the system. However, the \texttt{UELMAT} differs from the \texttt{UEL} in that it provides access to \texttt{ABAQUS}'s material library (\texttt{material\_lib\_mech}). We exploit this to obtain the stress tensor $\bm{\sigma}$ and the SMA material Jacobian $\mathcal{\bm{C}}$ for a given strain tensor. Abaqus2Matlab \cite{AES2017} is used to pre-process the input files. The code developed can be downloaded from \url{www.empaneda.com/codes}.  

\section{Results}
\label{Sec:FEMresults}

\begin{figure}[H]
  \makebox[\textwidth][c]{\includegraphics[width=1.1\textwidth]{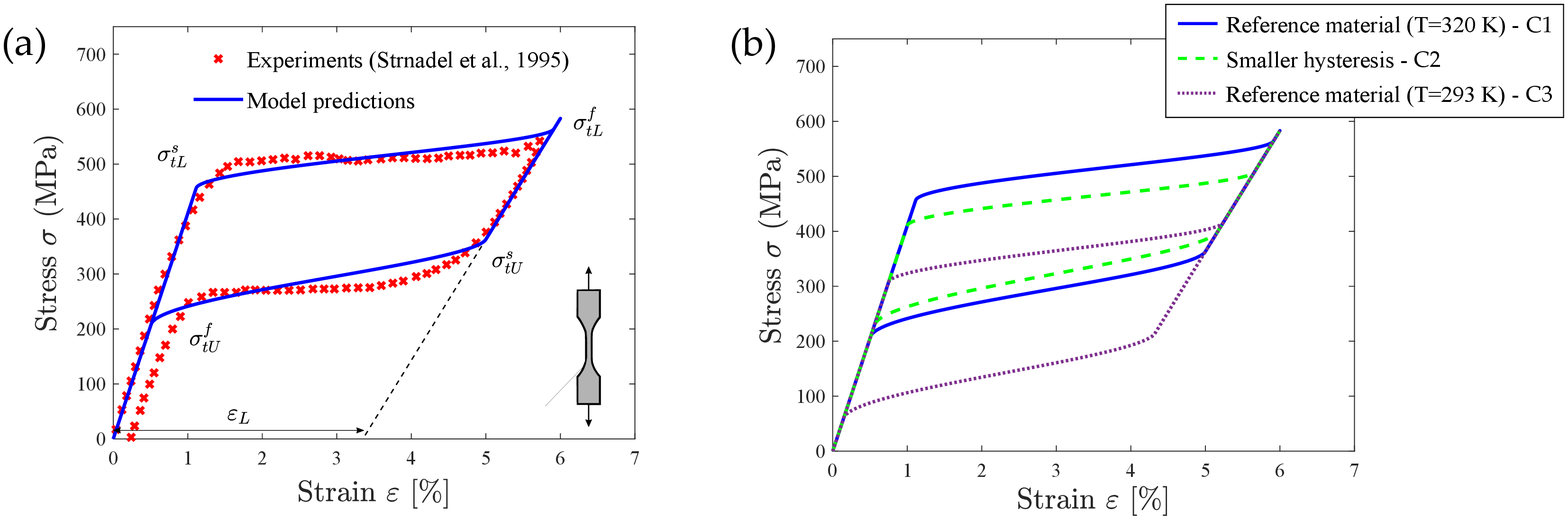}}%
  \caption{SMA constitutive behaviour: (a) reference material (C1), obtained by calibrating with the experiments by Strandel \textit{et al.} \cite{Strnadel1995} on an equiatomic NiTi; (b) three choices of material behaviour aimed at investigating the role of the stress-strain hysteresis (C2) and the temperature (C3).}
  \label{fig:SMASketch}
\end{figure}

The computational framework presented in Sections \ref{Sec:Theory} and \ref{Sec:NumModel} is used to gain insight into the fatigue behaviour of SMAs. To investigate the role of temperature and of the size of the stress-strain hysteresis, we conduct our numerical experiments considering, from the point of view of SMA deformation behaviour, three different scenarios; these are labelled C1, C2 and C3. The case C1 involves a reference material, which is determined by fitting the uniaxial stress-strain response of an equiatomic NiTi SMA, as measured by Strandel \textit{et al.} \cite{Strnadel1995} - see Fig. \ref{fig:SMASketch}a. Four regions can be readily identified. First, the response is linear elastic, and governed by the elastic constants of the austenite phase ($E_A$, $\nu_A$). Eventually, the stress reaches the threshold transformation stress for loading, $\sigma_{tL}^s$, and transformation to the martensite phase begins. The transformation process ends when the final transformation stress $\sigma_{tL}^f$ is reached. Upon further loading, the material behaviour is elastic but governed by the martensite elastic properties ($E_M$, $\nu_M$). If the load is removed, a martensite to austenite phase transformation will begin when the stress reaches the unloading threshold, $\sigma_{tU}^s$. The material will be fully austenitic when the final stress for transformation under unloading is attained, $\sigma_{tU}^f$. The values reported in Table \ref{tab:tab1length} provided the best fit to the experimental data by Strandel \textit{et al.} \cite{Strnadel1995} and are used throughout this paper unless otherwise stated. As in the experiments, numerical results for the reference material, C1, are obtained adopting a testing temperature of $T=320$ K. It is worth noting that, while the model seems to provide a good approximation to the uniaxial experiments by Strandel \textit{et al.} \cite{Strnadel1995} (Fig. \ref{fig:SMASketch}a), a better agreement can be obtained by incorporating additional modelling features. For example, the unified model of Lagoudas \cite{Lagoudas2008} includes additional smooth hardening parameters that enable following more closely the transformation regions. Also, the residual strain seen in the experiments after unloading can be readily captured by considering plastic deformation.

\begin{table}[H]
\caption{Material parameters used for the reference material, following the calibration with the uniaxial stress-strain measurements by Strnadel \textit{et al.} \cite{Strnadel1995} on an equiatomic nitinol SMA.}
\raggedleft
\centering
\begin{tabular}{l l} 
\hline
Parameter & Magnitude\\
\hline
Austenite's Young's modulus, $E_A$ (MPa) & 41000 \\
Martensite's Young's modulus, $E_M$ (MPa) & 22000\\
Austenite's Poisson's ratio, $\nu_A$ & 0.33 \\
Martensite's Poisson's ratio, $\nu_M$ & 0.33  \\
Transformation strain, $\varepsilon_L$ & 0.0335 \\
Start of transformation stress (Loading), $\sigma_{tL}^s$ (MPa) & 456.5 \\
End of transformation stress (Loading), $\sigma_{tL}^f$ (MPa) & 563.8 \\
Start of transformation stress (Unloading), $\sigma_{tU}^s$ (MPa) & 363 \\
End of transformation stress (Unloading), $\sigma_{tU}^f$ (MPa) & 209 \\
Martensite start temperature, $M_s$ (K) & 237\\
Martensite end temperature, $M_f$ (K) & 218\\
Austenite start temperature, $A_s$ (K) & 254\\
Austenite end temperature, $A_f$ (K) & 282\\
Reference temperature, $T_{ref}$ (K) & 320 \\
$\sigma$ vs $T$ slope (loading), $C_M|_{\sigma=300 \, \text{MPa}}$ (MPa/K) & 5.5 \\
$\sigma$ vs $T$ slope (unloading), $C_A|_{\sigma=300 \, \text{MPa}}$ (MPa/K) & 5.5\\
Material toughness $G_c$ (kJ/m$^2$) & 22.5 \\
\thickhline
\end{tabular}
\label{tab:tab1length}
\end{table}

\begin{figure}[H]
  \makebox[\textwidth][c]{\includegraphics[width=0.9\textwidth]{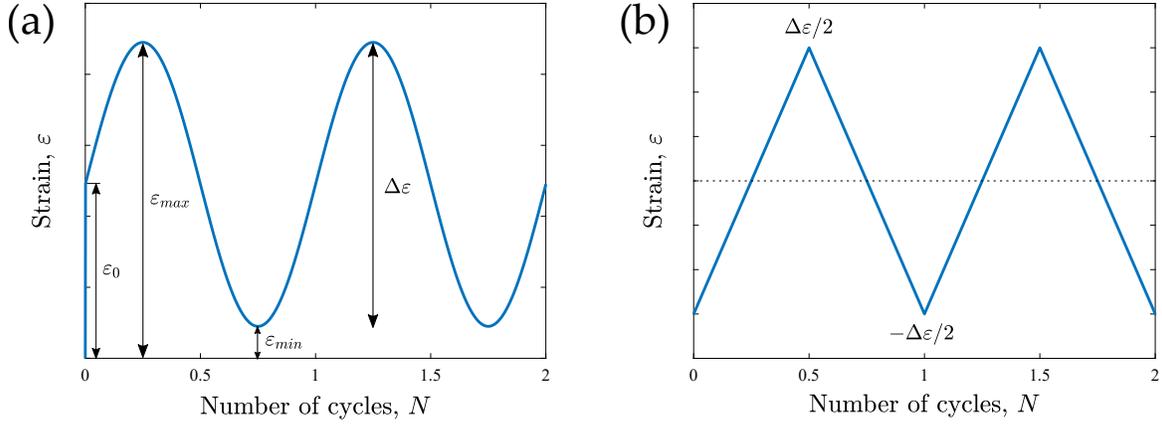}}%
  \caption{Cyclic loading histories employed: (a) sinusoidal and (b) piece-wise linear, shown here for a loading frequency of 1 Hz.}
  \label{fig:LoadingAmplitudes}
\end{figure}

The experimental literature reports a significant influence of the testing temperature and of the size of the stress-strain hysteresis (see, e.g., \cite{Wilkes2000} and Refs therein). To investigate the latter, we adopt the material parameters of Table \ref{tab:tab1length} but vary by 10\% the transformation stress thresholds, in what we refer to as case study C2. Specifically, we make the hysteresis loop smaller by reducing by 10\% the loading stress thresholds, such that $\sigma_{tL}^s=410.85$ MPa and $\sigma_{tL}^f=507.38$ MPa, and by increasing by 10\% the unloading stress thresholds, rendering $\sigma_{tU}^s=399.3$ MPa and $\sigma_{tU}^f=229.9$ MPa. To quantify the role of temperature, we use the material properties listed in Table \ref{tab:tab1length} but conduct our numerical experiments at $T=293$ K, this is referred to as C3. The three stress-strain responses resulting from these material choices are shown in Fig. \ref{fig:SMASketch}b. It can be seen that C2 results in a smaller hysteresis loop, relative to the reference material (C1), and that the reduction in temperature of C3 results in lower stress transformation thresholds. The same fracture properties are assumed for all cases; namely, a toughness of $G_c=22.5$ kJ/m$^2$, based on the data reported for NiTi \cite{Haghgouyan2018}, and a length scale of $\ell=0.145$ mm, which corresponds to a strength of 600 MPa according to Eq. (\ref{eq:AT2scec})a. Symmetric tension-compression behaviour is assumed but the focus will be on tensile behaviour; case studies where the load ratio $R$ is positive. As shown in Fig. \ref{fig:LoadingAmplitudes}, two types of cyclic loading histories will be used. The first one, Fig. \ref{fig:LoadingAmplitudes}a, is a sinusoidal loading history characterised by a maximum strain $\varepsilon_{max}$, a minimum strain $\varepsilon_{min}$, an initial strain amplitude $\varepsilon_0$ and a strain range $\Delta \varepsilon = \varepsilon_{max}-\varepsilon_{min}$. Considering also the definitions of load ratio $R=\varepsilon_{min}/\varepsilon_{max}$ and mean load $\varepsilon_m$, the sinusoidal loading history can be described as:
\begin{equation}
    \varepsilon = \varepsilon_m + \frac{\Delta \varepsilon}{2} \sin \left(  f 2  \pi t \right) = \frac{\Delta \varepsilon}{2} + R \frac{\Delta \varepsilon}{(1-R)} + \frac{\Delta \varepsilon}{2} \sin \left(  f 2  \pi t \right) \, ,
\end{equation}

\noindent where $f$ is the loading frequency and the strain amplitude is given by $\varepsilon_a=\varepsilon_{max}-\varepsilon_m$. Additionally, we will also consider a piece-wise linear variation of the applied displacement under constant amplitude, as shown in Fig. \ref{fig:LoadingAmplitudes}b. 

\subsection{Virtual $\Delta \varepsilon-N$ curves}

\begin{figure}[H]
  \makebox[\textwidth][c]{\includegraphics[width=1\textwidth]{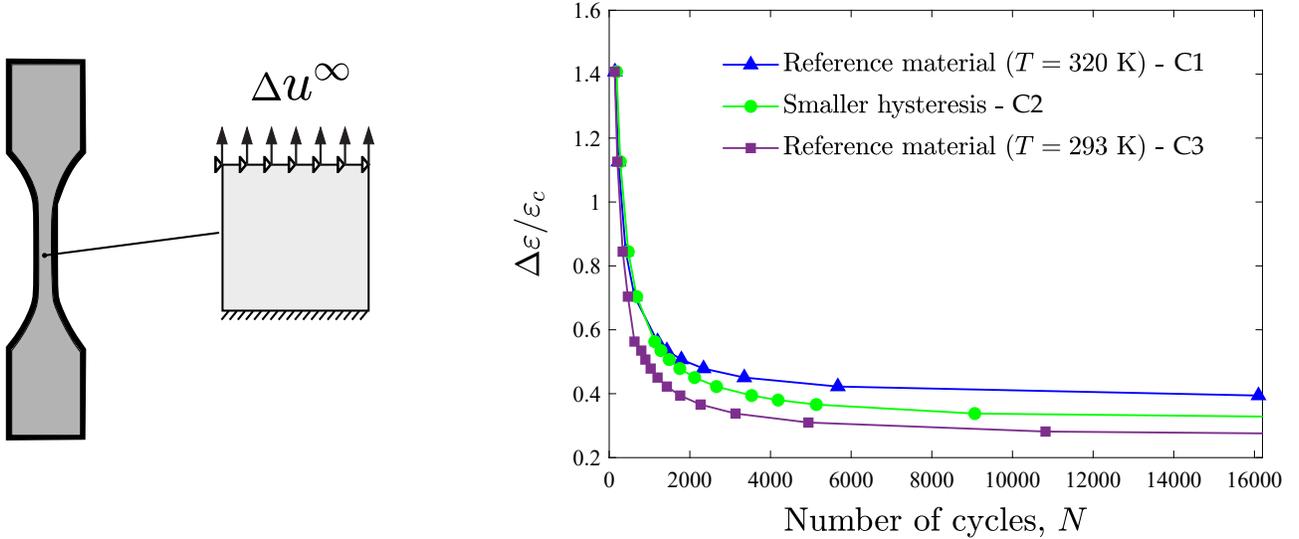}}%
  \caption{Virtual $\Delta \varepsilon-N$ curves for the three material behaviours considered. The strain range $\Delta \varepsilon$ is normalised by the critical strain, as given by Eq. (\ref{eq:AT2scec})b.}
  \label{fig:1D}
\end{figure}

We begin by conducting virtual uniaxial tests to estimate the load range versus number of cycles response, $\Delta \varepsilon-N$ curves. As shown in Fig. \ref{fig:1D}, we mimic the one-dimensional conditions of uniaxial testing by using a single square element and defining a tensile strain range $\Delta \varepsilon$ by prescribing the vertical displacement of the top edge $\Delta u^\infty$. The sinusoidal loading history shown in Fig. \ref{fig:LoadingAmplitudes}a is used, with $R=0.1$. Calculations are conducted using the phase field \texttt{AT2} model. The results obtained for the three material behaviours under consideration are shown in Fig. \ref{fig:1D}, with each marker corresponding to one virtual tensile experiment.

In all three cases, the results reveal a significant increase in the fatigue life with decreasing loading range, up to the point of appearing to show an endurance limit. The profile displayed agrees with that exhibited by experimental $\Delta \varepsilon - N$ curves \cite{Wilkes2000}. The three material behaviours considered exhibit a similar fatigue life for high loading ranges but differ significantly for low values of $\Delta \varepsilon$. For example, a strain range of $\Delta \varepsilon / \varepsilon_c=0.4$ results in a fatigue life of 16,100 cycles for the reference material (C1), while the number of cycles to failure is only 3,533 and 1,761 for the cases of C2 and C3, respectively. These results can be justified by the lower stress (and strain) thresholds for phase transformations of C2 and C3 relative to C1 - see Fig. \ref{fig:SMASketch}b. Inelastic transformation strains also accumulate in lower load levels and drive the evolution of fatigue damage. This is in agreement with experiments; samples that are initially in the austenitic phase exhibit shorter fatigue lives if the transformation stress is reduced \cite{Wilkes2000}. It has been suggested that increasing the transformation stress threshold will lead to an improved fatigue performance \cite{Hornbogen2004}. 

\subsection{Cracked square subjected to cyclic uniaxial loading}

\begin{figure}[H]
  \makebox[\textwidth][c]{\includegraphics[width=1.1\textwidth]{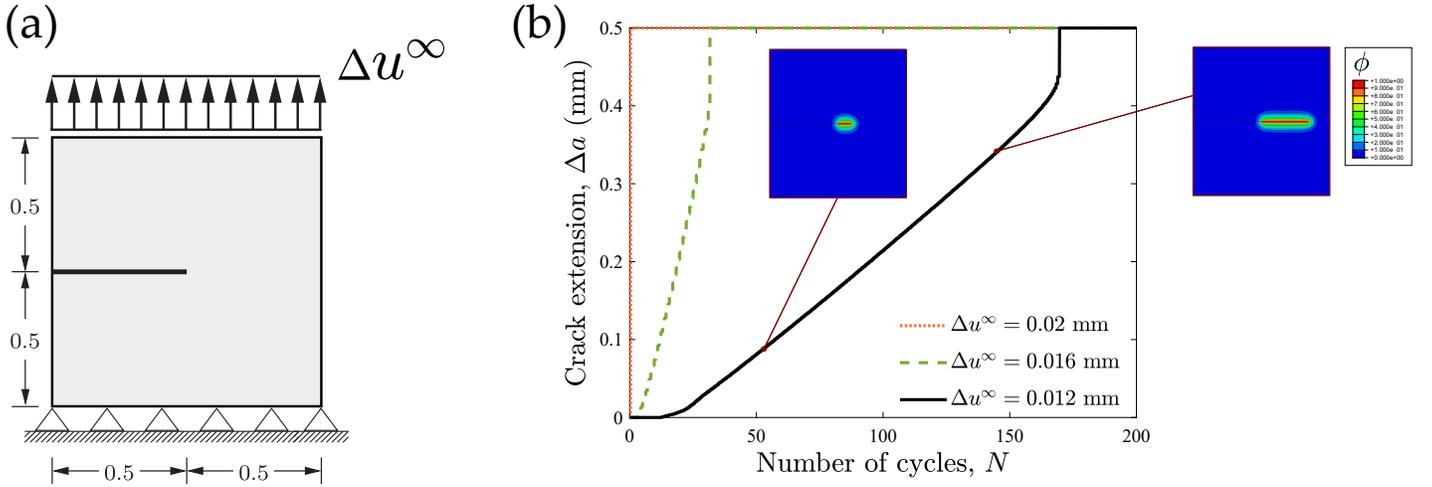}}%
  \caption{Cracked square plate: (a) dimensions (in mm) and loading configuration, and (b) crack extension versus number of cycles results for varying load ranges. The results have been obtained with the \texttt{AT1} model.}
  \label{fig:SENT1}
\end{figure}

The second case study deals with a paradigmatic benchmark in phase field fracture: crack propagation in a square plate subjected to tension, see Fig. \ref{fig:SENT1}a. As in Refs. \cite{Carrara2020,TAFM2020,Golahmar2022}, we prescribe the remote load in a piece-wise linear manner. The material properties of Table \ref{tab:tab1length} are adopted in this example; i.e., we use C1, the reference material, and assume a test temperature of $T=320$ K. The focus is on assessing the influence of the constitutive choice for the crack density function: \texttt{AT1} vs \texttt{AT2} models. The geometry, dimensions and loading configuration are given in Fig. \ref{fig:SENT1}a. An initial horizontal crack is introduced geometrically, which goes from the left side of the specimen to its centre. The domain is discretised with approximately 23,000 linear quadrilateral elements, with the characteristic element length in the crack growth region being two orders of magnitude smaller than the phase field length scale, which is more than sufficient to ensure mesh convergence \cite{CMAME2018,Pandolfi2021}. The results obtained for different load amplitudes and the \texttt{AT1} model are shown in Fig. \ref{fig:SENT1}b. In agreement with expectations, we observe a faster crack growth rate with increasing $\Delta u^\infty$. The crack propagates in a stable manner, following the expected mode I crack path, as in the case of static loading \cite{CMAME2021}. 

\begin{figure}[H]
  \makebox[\textwidth][c]{\includegraphics[width=0.9\textwidth]{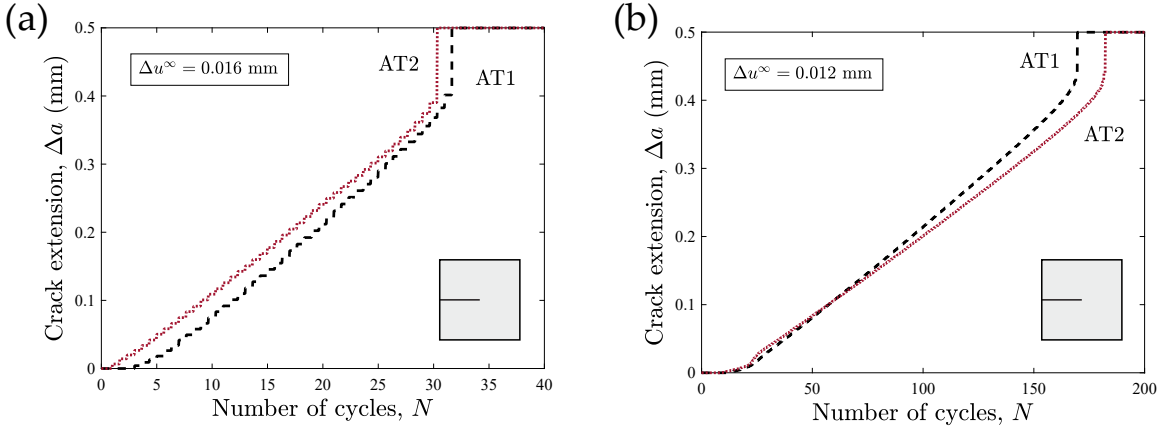}}%
  \caption{Cracked square plate. Influence of the constitutive choice for the crack density function, \texttt{AT1} \cite{Pham2011} versus \texttt{AT2} \cite{Bourdin2000} formulations: crack extension versus number of cycles for load ranges (a) $\Delta u^\infty=0.016$ mm, and (b) $\Delta u^\infty=0.012$ mm.}
  \label{fig:SENT2}
\end{figure}

We proceed to compare the predictions obtained with the \texttt{AT2} and \texttt{AT1} phase field models. The same magnitude of $G_c$ and $\ell$ is assumed in both cases, which leads to different values of the critical stress and the critical strain - see Eqs. (\ref{eq:AT1scec})-(\ref{eq:AT2scec}). As shown in Fig. \ref{fig:SENT2}, the predictions obtained with both models show a rather close agreement but their ranking, in terms of fatigue life, is sensitive to the load range. First, for both $\Delta u^\infty=0.016$ mm and 0.012 mm, we can observe that the initiation of crack growth takes place earlier for the \texttt{AT2} model. This is attributed to the presence of a damage threshold in the \texttt{AT1} model and to the higher material strength that is obtained for \texttt{AT1} if the same $G_c$ and $\ell$ are sampled in Eqs. (\ref{eq:AT1scec})-(\ref{eq:AT2scec}). A similar observation has been recently made in the context of fatigue crack growth in non-linear kinematic and isotropic hardening elastic-plastic solids \cite{CMAME2022}. However, in both Figs. \ref{fig:SENT2}a and \ref{fig:SENT2}b, fatigue crack growth rates appear to be larger for the \texttt{AT2} model, bringing predictions closer together, in the case of $\Delta u^\infty=0.016$ mm, and even resulting in a shorter fatigue life for the case of $\Delta u^\infty=0.012$ mm. It should be noted that, for a given $\ell$ and $G_c$, the critical strain is larger for \texttt{AT1} relative to the \texttt{AT2} case - see Eqs. (\ref{eq:AT1scec})-(\ref{eq:AT2scec}).

\subsection{Boundary layer model: Paris law behaviour}

\begin{figure}[H]
  \makebox[\textwidth][c]{\includegraphics[width=1\textwidth]{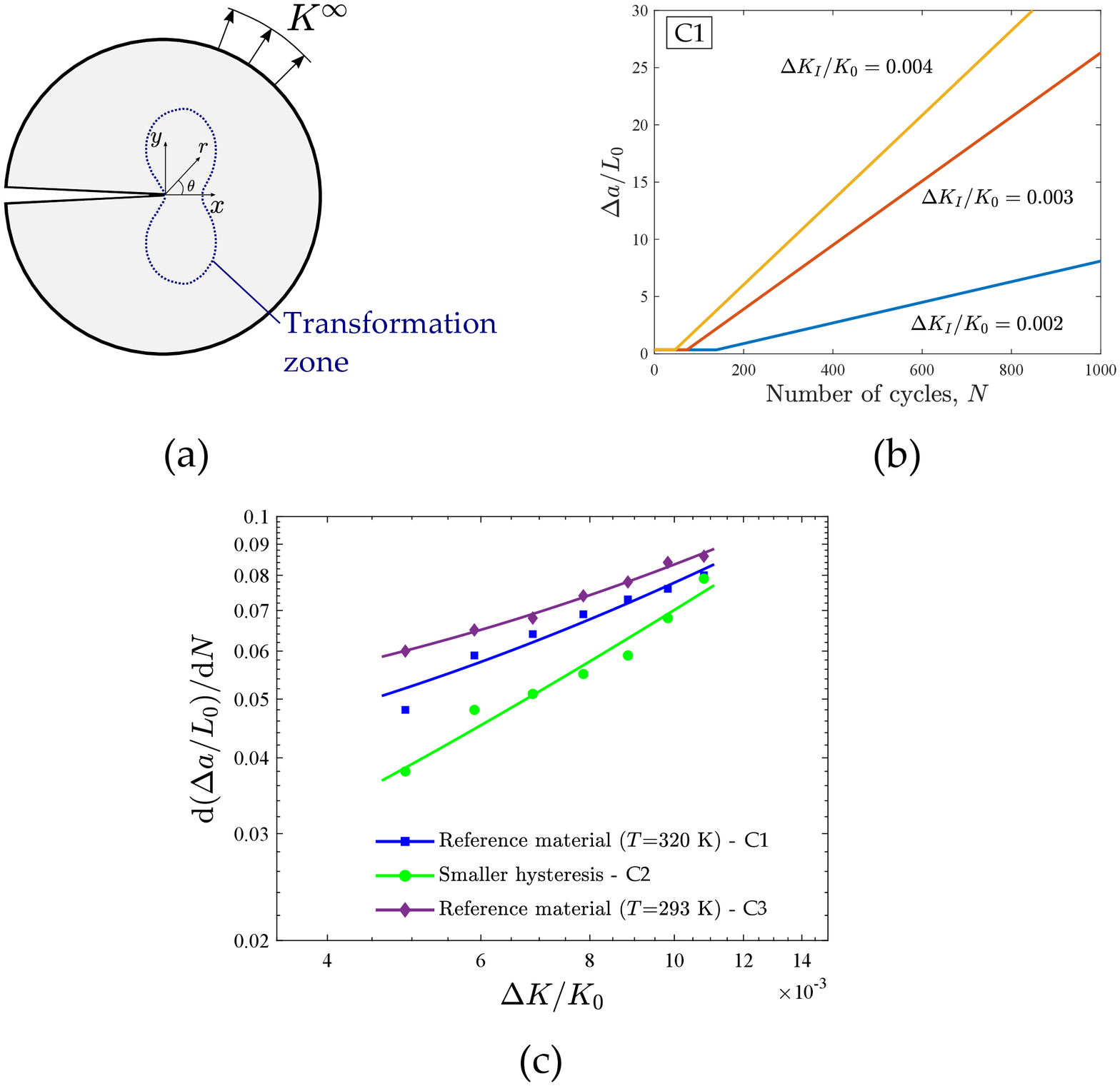}}%
  \caption{Boundary layer model, crack growth rates in the Paris law regime: (a) sketch of the boundary layer configuration and the small transformation zone assumption, (b) crack extension versus number of cycles for various loading ranges, and (c) fatigue crack growth rate versus load range for the three material behaviours considered.}
  \label{fig:BLresults}
\end{figure}

Crack growth rates under small transformation zone conditions are quantified to assess the influence of testing temperature and hysteresis shape on the Paris law regime. Here, it should be emphasised that Paris law curves are an outcome of the model; a prediction and not an input. As outlined in Fig. \ref{fig:BLresults}a, a boundary layer model is used, whereby a remote elastic $K_I$ field is prescribed by defining the displacement of the nodes located in the outer boundary in agreement with William's \cite{Williams1957} solution. Accordingly, the horizontal and vertical displacements in the outer nodes are respectively equal to:
\begin{equation}
    u_x = K_I \frac{1+\nu_A}{E_A} \sqrt{\frac{r}{2 \pi}} \left( 3- 4 \nu_A - \cos \theta \right) \cos \left( \frac{\theta}{2} \right)
\end{equation}
\begin{equation}
    u_y = K_I \frac{1+\nu_A}{E_A} \sqrt{\frac{r}{2 \pi}} \left( 3- 4 \nu_A - \cos \theta \right) \sin \left( \frac{\theta}{2} \right)
\end{equation}

\noindent where $r$ and $\theta$ are the coordinates of a polar coordinate system centred at the crack tip. For normalisation purposes, a reference stress intensity factor can be defined as:
\begin{equation}
    K_0 = \sqrt{\frac{E_A G_c}{(1-\nu_A^2)}} \, ,
\end{equation}

\noindent together with the following fracture process zone length:
\begin{equation}
    L_0 = \frac{G_c \left( 1- \nu_A^2 \right)}{E_A} \, .
\end{equation}

The model uses a total of 42,081 degrees-of-freedom (DOF) and the characteristic element length along the crack propagation region is taken to be more than 10 times smaller than the phase field length scale. The finite element results obtained for the three material behaviour case studies considered are shown in Fig. \ref{fig:BLresults}. First, curves of crack extension ($\Delta a/L_0$) versus number of cycles ($N$) are obtained for various loading amplitudes (Fig. \ref{fig:BLresults}b). The slope of the curve is then measured and plotted against the load range ($\Delta K/K_0$) in a log-log plot (Fig. \ref{fig:BLresults}c). It can be seen that, in agreement with expectations, the model predicts increasing fatigue crack growth rates with rising load amplitude. 

The results obtained reveal several interesting patterns. First, differences are relatively small between the three cases. A fit to the well-known Paris law expression,
\begin{equation}
    \frac{da}{dN} = C \Delta K^m \, ,
\end{equation}
\noindent gives values of $m$ ranging from 1.2 to 1.4. A small influence of the martensitic transformation on Paris law parameters has also been observed experimentally but slightly larger values have been reported for the Paris law coefficient $m$ in NiTi; from $m=2.2$ to $m=3$, depending on the testing environment and alloy composition \cite{Wilkes2000,Sgambitterra2019,Haghgouyan2021}. Second, faster crack growth rates are observed when decreasing the temperature, in agreement with the susceptibility trends observed for the uniaxial experiments (see Fig. \ref{fig:1D}). This is also in agreement with experimental testing - the higher the temperature the higher the cycles to failure and the lower the crack growth rate \cite{Sgambitterra2019}. Finally, the smaller hysteresis material (C2) exhibits lower crack growth rates, particularly for small $\Delta K$ values. This appears to contradict the results obtained under uniaxial tension, where the reference material exhibited the highest resistance to fatigue damage. Smaller hysteresis have been related to longer fatigue lives and thus are particularly suited for fast response actuators and precision control applications in the aerospace sector \cite{Lagoudas2009}. 

\subsection{Fatigue failure of a 3D lattice}

\begin{figure}[H]
  \makebox[\textwidth][c]{\includegraphics[width=1\textwidth]{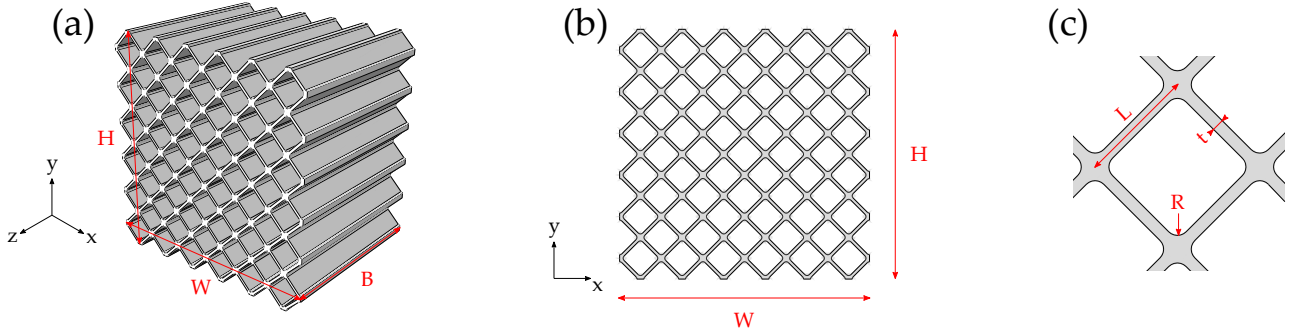}}%
  \caption{Geometry of the 3D lattice: (a) isometric view, (b) front view, and (c) details of the individual struts.}
  \label{fig:LatticeSketch}
\end{figure}

Finally, we demonstrate the capabilities of the modelling framework in predicting fatigue failures in complex geometries at large scales. In particular, we simulate the failure of a diamond lattice structure under repeated compression cycles \cite{Cote2006}. The geometry of the diamond lattice material simulated is shown in Fig. \ref{fig:LatticeSketch} through its isometric and in-plane views. The height and the width equal $H=W=18$ mm, while the depth equals $B=13$ mm. Each individual strut has a length of $L=2.09$ mm and a thickness of $t=0.2$ mm. A fillet radius of $R=0.2$ mm is introduced to suppress stress singularities. 

\begin{figure}[H]
  \makebox[\textwidth][c]{\includegraphics[width=1.1\textwidth]{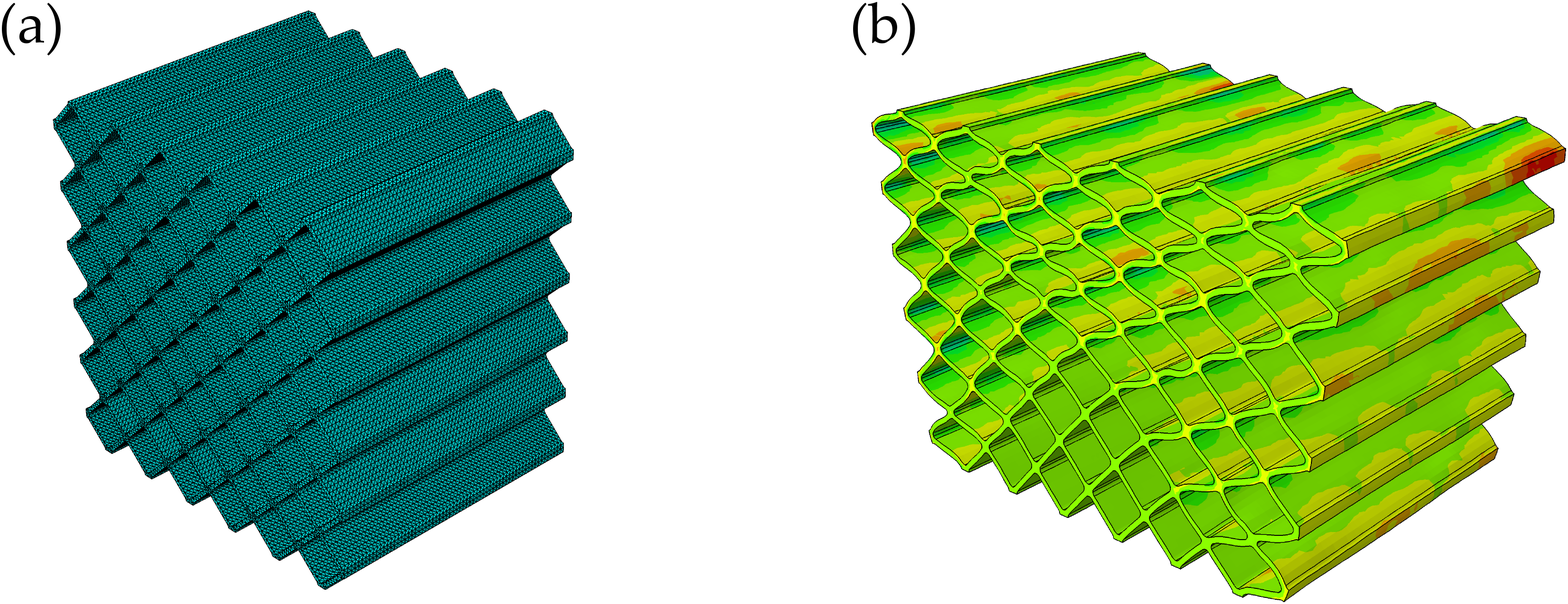}}%
  \caption{3D lattice: (a) finite element mesh and (b) deformed shape at $u_{min}$ with a scale factor of 10.}
  \label{fig:LatticeMesh}
\end{figure}

The entire geometry is modelled to showcase the efficiency of the computational framework in dealing with large scale problems. The domain is discretised with 10-node quadratic tetrahedral elements. As shown in Fig. \ref{fig:LatticeMesh}a, the mesh is uniform and the characteristic element length equals 0.2 mm, which is 5 times smaller than the phase field length, which is taken to be equal to $\ell=1$ mm in this case study. All the other properties are those listed in Table \ref{tab:tab1length} (the reference material) and the phase field \texttt{AT2} model is used. The model contains more than 7 million DOFs and is run in parallel, using 16 cores. The diamond lattice has its three displacement components constrained at the bottom ($y=0$) and is subjected to a vertical ($u_y$) displacement at the top. The cyclic load is applied using a piece-wise linear amplitude (Fig. \ref{fig:LoadingAmplitudes}b), with a load range of 0.2 mm, with $u_{max}=0$ and $u_{min}=-0.2$ mm. The deformed shape of the lattice undergoing compression is shown in Fig. \ref{fig:LatticeMesh}b, upon the application of a scaling factor of 10. It can be observed that, as expected, the struts located near the centre of the top side of the lattice undergo the largest deformations.

\begin{figure}[H]
  \makebox[\textwidth][c]{\includegraphics[width=1.3\textwidth]{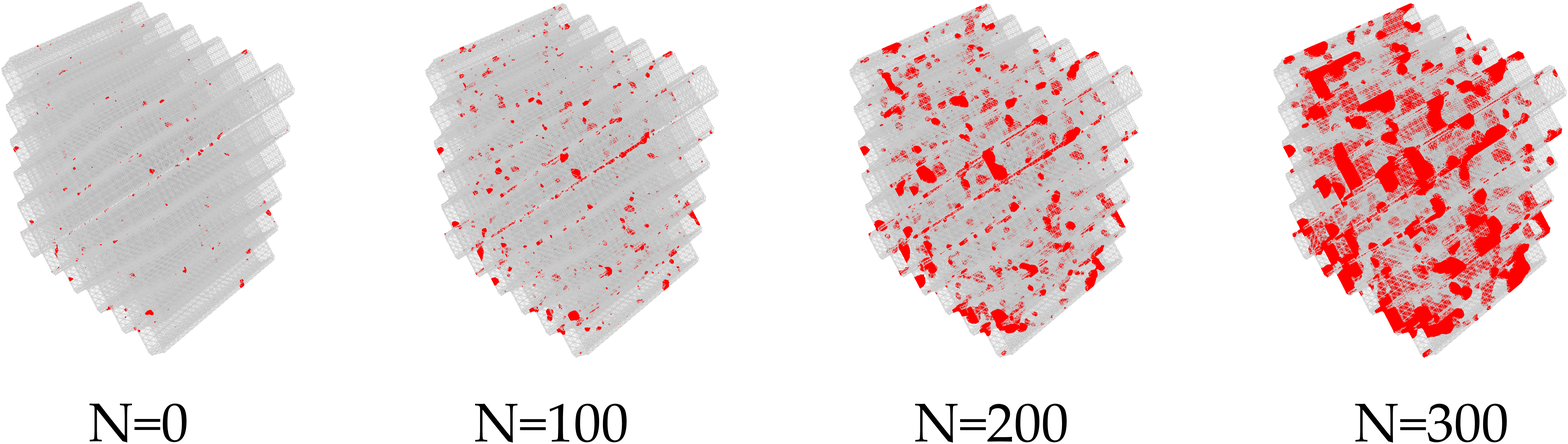}}%
  \caption{3D lattice. Contours of defects (regions with $\phi>0.95$) as a function of the number of loading cycles. Results are shown in the undeformed state.}
  \label{fig:LatticeResults}
\end{figure}

An initial distribution of defects is randomly assigned to account for the role that initial cracks and voids can play in the fatigue failure process. This is particularly relevant for these lattice structures as they are typically fabricated using additive manufacturing. The initial distribution of defects is shown in Fig. \ref{fig:LatticeResults}a, where the areas with $\phi>0.95$ have been plotted before applying mechanical load. Figs. \ref{fig:LatticeResults}b-d show the evolution of defects in time. It can be observed that the initial defects grow and eventually merge with neighboring defects, leading to the nucleation of cracks of significant size and the loss of load carrying capacity. The capabilities of the present computational framework in capturing complex cracking phenomena are demonstrated. 

\section{Conclusions}
\label{Sec:ConcludingRemarks}

We have presented a new formulation for predicting fatigue crack growth in Shape Memory Alloys (SMAs). The model is based on a phase field description of the crack-solid interface and accommodates the two main constitutive choices for the crack density function; the so-called \texttt{AT1} and \texttt{AT2} approaches. This phase field description has been enriched with a fatigue degradation function and a fatigue history variable. Damage, static and cyclic, is driven by the total strain energy density, including contributions from both elastic and transformation mechanical fields. Material deformation is captured by the constitutive model for SMAs developed by Auricchio and co-workers \cite{Auricchio1997}, where a Drucker-Prager type of transformation surface is used. The theoretical framework is numerically implemented using the finite element method, and a robust monolithic quasi-Newton solution scheme is used to significantly reduce the computational cost. The computational model is used to simulate fatigue failures in multiple configurations. Several loading histories, sinusoidal and piece-wise linear, are adopted. Three case studies are considered for the material parameters with the aim of gaining insight into the role of transformation stress thresholds, temperature and shape of the hysteresis zone. Our simulations address: (i) the $\Delta \varepsilon$ vs number of cycles to failure response of smooth samples undergoing uniaxial tension, (ii) the growth of fatigue cracks in single edge notched specimens, (iii) the prediction of fatigue crack growth rates to quantify Paris law parameters, and (iv) the fatigue failure of a 3D diamond lattice structure. The results are mechanistically interpreted and discussed in the context of the experimental fatigue SMA literature. The main findings are:

\begin{itemize}
    \item The model adequately captures the fatigue behaviour of SMAs, including the sensitivity to temperature and to the stress-strain hysteresis size of fatigue crack growth rates and the number of cycles to failure.
    
    \item The comparison between \texttt{AT1} and \texttt{AT2} phase field models reveals that the initiation of growth occurs first for \texttt{AT2} but that higher fatigue crack growth rates are predicted by the \texttt{AT1} formulation.
    
    \item The phase transformation process has a limited effect on Paris law coefficients, in agreement with experimental observations. However, the $m$ values attained are slightly below those measured in the literature.
    
    \item Complex fatigue cracking phenomena can be captured efficiently and without convergence problems. These include the growth and coalescence of multiple defects in arbitrary geometries and dimensions.
\end{itemize}

The framework developed can be used to optimise the design of SMA components and map safe regimes of operation. Potential avenues for future work include the modelling of thermal fatigue and extending the model to account for plasticity.

\section*{Acknowledgements}

M. Simoes acknowledges financial support from the EPSRC (grant EP/R512461/1) and from the ESA (Contract no. 4000125861). E. Mart\'{\i}nez-Pa\~neda was supported by an UKRI Future Leaders Fellowship (grant MR/V024124/1).






\bibliographystyle{elsarticle-num}
\bibliography{library}

\end{document}